\documentclass[namedreferences]{solarphysics}
\usepackage[showbiblabels]{spr-sola-addons}

\usepackage{mathtools}

\usepackage{fourier}
\usepackage[pdftex]{graphicx}

\usepackage{gensymb}

\usepackage{breakurl}        
\usepackage[table]{xcolor}

\definecolor{darkblue}{RGB}{0, 0, 120}
\definecolor{darkred}{RGB}{80, 0, 0}
\usepackage[colorlinks = true,
            linkcolor = darkblue,
            urlcolor  = darkblue,
            citecolor = darkred,
            anchorcolor = darkblue]{hyperref}

\chardef\us=`\_
\begin{document}
\begin{article}
\begin{opening}

\title{A modern reconstruction of  Richard Carrington’s observations (1853-1861)
\\ {\it Solar Physics}}

\author[addressref={aff1,aff4},corref,email={shreya.bhattcharya@oma.be}]{\inits{B.}\fnm{Bhattacharya }~\lnm{S.}}
\author[addressref=aff2,email={teague@cantab.net}]{\inits{T.}\fnm{Teague}~\lnm{E.T.H.}}
\author[addressref={aff1,aff3},email={stephen.fay@mail.mcgill.ca}]{\inits{F.}\fnm{Fay}~\lnm{S.}}
\author[addressref=aff1,email={laure.lefevre@oma.be}]{\inits{L.}\fnm{ Lefèvre}~\lnm{L.}}
\author[addressref=aff4,email={maarten.jansen@ulb.ac.be}]{\inits{J.}\fnm{Jansen}~\lnm{M.}}\author[addressref=aff1,email={frederic.clette@oma.be}]{\inits{C.}\fnm{Clette}~\lnm{F.}}
\address[id=aff1]{ Royal Observatory of Belgium, WDC-SILSO}
\address[id=aff2]{WDC-SILSO observer, UK}
\address[id=aff3]{McGill University, Montreal, CA}
\address[id=aff4]{Universit\'e Libre de Bruxelles, Bruxelles, Belgium}

\runningauthor{Bhattacharya\textit{ et al}.}
\runningtitle{A modern reconstruction of Carrington’s observations }

\begin{abstract}
The focus of this article is a re-count of Richard Carrington's original sunspot observations from his book drawings \citep{CarringtonBook1863} by an observer from the World Data Center-SILSO (WDC-SILSO, \url{http://www.sidc.be/silso/home}) network, Thomas H. Teague (UK). This modern re-count will enable the  use of Carrington's observations in the recomputation of the entire Sunspot Number series in a way Carrington's original counts \citep{2014SoPh..289...79C} did not. Here we present comparison studies of the new re-counted series with contemporary observations, new data extracted from the Journals of the Zurich Observatory and other sources of Carrington's own observations and conclude that Carrington's group counting is very close to the modern way of counting while his method for counting individual spots lags significantly behind modern counts. We also test the quality and robustness of the new re-count  with methods developed in \cite{2019ApJ...886....7M}.
\end{abstract}
\keywords{Sun, sunspots, sunspot number}
\end{opening}
\section{Introduction}
 \label{S-Introduction} 

In 1843, R. Wolf founded a journal called the "Mittheilungen der Naturforschenden Gesellschaft in Berne". From 1848 \citep{1848nvds.book.....W} until his death \citep{1894MiZur...9..109W}, he published a yearly book with all of his findings, including sunspot observations as far back as Galileo \citep{1861MiZur...2...83W}, and the sunspot numbers collected by him, his European colleagues and his auxilliary observers in this journal: in this work, they will be consistently referred to as "the Mittheilungen".

Between 2017 and 2019, the Royal Observatory of Belgium (\url{https://www.astro.oma.be/en/}), and more precisely the WDC-SILSO, conducted a program to digitize all the data contained in the published Mittheilungen. After a community-wide effort that led to a fully recalibrated series for the sunspot number (SN) and a reconstruction of the group number (GN) in 2015 \citep{2016SoPh..291.2479C}, the next step is to reconstruct the sunspot number series from the raw original data. A large part of the raw data can be found in the Mittheilungen, although there remain a few caveats that we'll explain along the way.\\

Richard Carrington's observations, that span 1853 to 1861, overlap a critical period near the beginning of the series started by R. Wolf in 1849 \citep{1848nvds.book.....W}. However the data R. Wolf reported in the Mittheilungen \citep{1865MiZur...2..193W,1874MiZur...4..173W} concerning Carrington's observations are incomplete and inhomogeneous (see section \ref{sec:Carrington data in mitt}). This is why , Thomas H. Teague, an active observer of the WDC-SILSO network recently studied Carrington's observations \citep{1996JBAA..106...82T} and made a re-count of Carrington’s sunspot observations with modern methods  used by the observers of the WDC-SILSO network today. This re-count opens up the possibility of comparing stable observers straddling his observations including Rudolf Wolf himself.\\

Thanks to the work by Thomas H. Teague, and the recent digitization of the Mittheilungen we will now be able to realize a first quality assessment of the early part of the Sunspot Number series. Section \ref{S-aug} presents Richard Carrington's  and his contemporaries' observations. Section \ref{sec:data sources} presents all the data sources used in this article. Section \ref{sec:modern reconstruction} introduces the modern reconstruction of sunspot numbers by Teague while section \ref{s-com} compares  the different reconstructions of Carrington's original data. Section \ref{sec:comparison obs} focuses on comparisons with overlapping long-term observers and assesses the quality of this reconstruction with modern methods. Section \ref{sec:conclusions} presents our conclusions.


\section{Sunspot observations in the mid-nineteenth century}
\label{S-aug}

\subsection{Richard Carrington's observations}
Between November 1853 and March 1861, the English astronomer Richard Carrington (1826–1875) made more than 5,000 observations of nearly 1,000 sunspot groups from his private observatory at Redhill in Surrey (UK). Using a 4.5-inch refractor and applying his own method of deriving heliographic coordinates from timed transits, he measured the positions of the groups he observed with unprecedented accuracy. \cite{2014SoPh..289...79C} have concluded that Carrington’s data were of high quality. H. W. Newton (\citealp{1958fasu.book.....N}, p.37) considered Carrington's sunspot measurements comparable in accuracy with results obtained a century later with solar photographs in a measuring machine.\\

Richard Carrington projected the image of the solar disc to a diameter of 11 inches “to allow for unintentional exaggeration” (\citealp{CarringtonBook1863}, p. 9) when drawing to a scale of 12 inches. For each day on which he observed, he produced a whole disc drawing in a bound logbook. According to his own explanation, he began by drawing individual groups, indicating by letters of the alphabet “the particular nuclei or points of the nuclei selected for observation” (\citealp{CarringtonBook1863}, p. 9). These he then measured using his transit method of determining sunspot positions. An inspection of selected whole disc drawings \citep{1978MmRAS..85....1B} suggests that his practice was to use the positions thus derived to plot each measured group or spot correctly on the day’s drawing before completing the remaining details by eye.

In 1858, the death of his father forced Carrington to take over the family brewing business, significantly disrupting his scientific activities. In order to continue his sunspot observation he had to employ a succession of assistants (\citealp{CarringtonBook1863}, p. 3):
\textit{During 1854, 1855, 1856, and 1857, I took all the observations myself, but was aided in their reduction by my assistant Mr. Simmonds. During 1858, I had no assistant, and my arrangements were greatly disturbed by the sudden death of my father, the superintendence and ultimate taking up of whose affairs caused much absence from Redhill. In 1859, I had the assistance for two short intervals, of Mr. J. Breen and Mr. H. Criswick, and towards the close of the year engaged Dr. von Bose, who shortly made himself familiar with my methods of observation and reduction, and applied them with much success through the year 1860, when the spots were very numerous and complicated. On his departure, I engaged Dr. Schroeder then in Paris on leave of absence by the department of public education of Hanover, and who observed with tolerable success for about three months.} It was the premature departure of Dr Schroeder in 1861 March that finally compelled Carrington to wind up his series of solar observations.

When the time came {\bf for publication}, Carrington adopted a different method of presenting his results. Instead of reproducing his daily whole disc drawings in his published volume, he decided to provide {\bf two sets of illustrations}: first, a series of {\bf rotation drawings} showing each group in its position “in its most typical aspect” (\citealp{CarringtonBook1863}, p. 16); second, a series of {\bf sequences of detailed individual drawings}, arranged vertically, in which he recorded the successive daily appearance of each group that he had observed more than once during its passage across the Sun’s face (\citealp{CarringtonBook1863}, p. 3). In addition, he assigned serial numbers to the groups he had observed, and tabulated their heliographic coordinates from day to day. Finally, he appended a section of notes in which he commented upon groups of particular interest. This means that {\bf the work presented here is based on rotation drawings and sequences of detailed drawings not on daily whole disc drawings} that do \textit{not} appear in \cite{CarringtonBook1863}.\\

Although Carrington was not personally responsible for making all of the contemporaneous whole disc drawings, he himself undertook the preparation of the rotation charts and sequences of detailed drawings for inclusion in his published volume (\citealp{CarringtonBook1863}). In performing that task, he reproduced the sequences of detailed drawings on a rectilinear grid of half-inch (1.27cm) squares at a uniform scale of 0.5” = 10 \degree, thereby compensating for any foreshortening. As with the original disc drawings, he first plotted the points whose positions he had measured, interpolating the remaining details of each group referencing to the whole disc drawings.
\subsection{Contemporary Observations}\label{sec:contemporary observations of carrington}
The mid-nineteenth century marks the official start of the consistent observations (daily) providing the International Sunspot Number. Early on, Prof. R. Wolf did not necessarily distinguish his observations from those of other observers' he used, although he mentioned their names in the Mittheilungen. Over the same period as Carrington (1825-1867), Samuel Heinrich Schwabe observed the Sun from Dessau, Germany \citep{2011AN....332..805A,2013MNRAS.433.3165A}.  Schwabe was the reference observer for the start of the sunspot series.

\begin{figure}[hbt!]   
\centerline{\includegraphics[width=12cm]{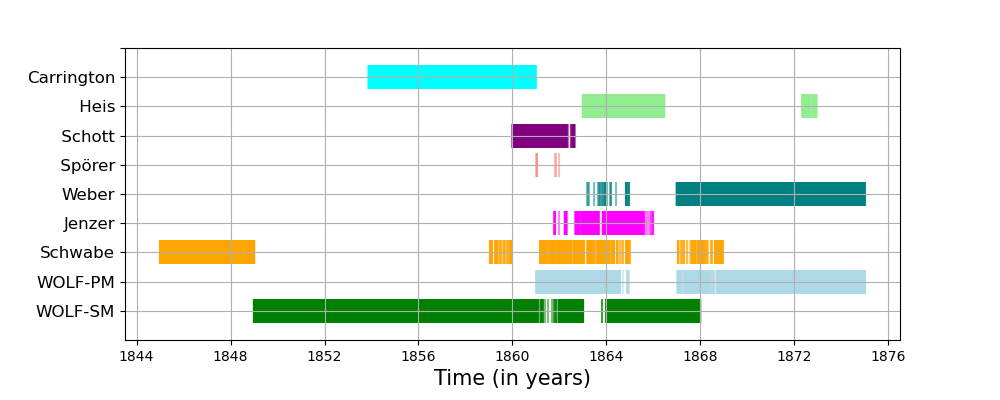}}
\caption{Observers contemporary to Richard Carrington as found in the Mittheilungen journals. }
\label{f-1}
 \end{figure}

Figure \ref{f-1} shows the observers contemporary to Carrington as they appear in the digital version of the Mittheilungen journals of Zurich 

The most prominent observer in figure \ref{f-1} is {\bf Pr. Rudolf Wolf}, who observed from Zurich with two main instruments between 1849 and 1893 \citep{2016SoPh..291.2505F}. Figure \ref{f-1} does not extend all the way to 1893 because we are looking specifically  only on the period where Wolf and Carrington overlap. Note that Schwabe continues to observe during 1849-1859 but his data is used by Wolf only to fill gaps in Wolf's own data and are referenced as Wolf's in the Mitteilungen. So in this period Wolf SM data is actually a mix Wolf, Schwabe and several other observers.

Friedrich Wilhelm Gustav {\bf Spörer} (1822–1895) started taking positional data of sunspots in December 1860 after being inspired by the observations of Richard Carrington. He became one of the first three astronomers who were supposed to study the physical processes on the Sun and stars in the Astrophysical Observatory Potsdam, which was founded in 1874 \citep{2020LRSP...17....1A}. His observations (1861-1894) were digitized by \cite{2015AN....336...53D}, but do not overlap directly with Carrington's. 

Heinrich {\bf Weber} observed from 1859-1883 in Peckeloh (40km east of Münster, Germany) and Eudard {\bf Heis} observed from 1863-1866 and again in 1872 from M{\"u}nster (Germany), and they both decided to share their observations with Prof. Wolf in 1863, observations done in the same manner that Wolf introduced so that he could use them as auxiliary observers \citep{1864MiZur...2..163W}.\\

Charles Anthony {\bf Schott} was the superintendent of the Coast Survey (Washington D.C., US) and reported his observations to Wolf for the years 1860 and 1861. Wolf mentioned \citep{1864MiZur...2..163W} that the observations were made following the exact procedure introduced by Wolf, but they needed to be scaled by a factor of 1.16 to bring them to the same scale as his own. Wolf used these observations to fill gaps in his own observation for the years 1860 and 1861. Some of his original drawings can be found at \url{https://www.ngdc.noaa.gov/stp/space-weather/solar-data/solar-imagery/photosphere/sunspot-drawings/charles-schott/}.\\

Emil {\bf Jenzer} was Wolf's pupil and observed from Bern from 1861 to 1865 \citep{1864MiZur...2..163W, 2016SoPh..291.2505F} and sported a scaling factor of 0.85 versus R. Wolf's observations. The refractor had an aperture of 83 mm and a focal length of 1300 mm.\\


In addition to the sources found in the Mittheilungen, Christian H. F. {\bf Peters} observed the Sun from the Hamilton Observatory, in Clinton, New York (US) over the period 1860-1870 and his observations can be found in \cite{2014SoPh..289...79C}. \\

In the work presented below, we are only going to use datasets that completely overlap Carrington's observations. This means we are going to use the Carrington data extracted by \cite{2014SoPh..289...79C}, with overlapping data from Schwabe \cite{2013MNRAS.433.3165A}  as well as Wolf's and Carrington's data extracted from the Mittheilungen.

\subsection{Scaling factors}
To scale the spots, groups or Wolf counts of an observer ($i$) to the primary observer, we use {\bf scaling factors} often called {\bf k-factors or k-coefficients}. Wolf counts (W) are a combination of groups and spots counts according to the definition of Wolf \citep{1848nvds.book.....W} following the \textbf{formula} $W=10g+f$, where $g$ is the number of groups and $f$ the number of spots \citep{1985Izenman}. Pr. Rudolf Wolf introduced k-factors when defining the original Sunspot Number to bring the observations of other observers to his own observations' scale. The modern day production of the International Sunspot Number (SN) by the WDC-SILSO follows a similar approach; it scales the values obtained from station $i$ to the values of a pilot station by multiplying their numbers by a k-factor, given as:
\begin{equation}
k_i(t) = \frac{pilot(t)}{Y_i(t)} 
\end{equation}
where $Y_i(t)$ are the Wolf Counts of station i, observed at time t and pilot(t) is the value of the pilot station explained in \cite{2007AdSpR..40..919C} and \cite{2019ApJ...886....7M}. \\

Note that the application of {\bf k-factors} is a historical heritage that needs re-evaluation with modern means in the context of the reconstruction of the Sunspot Number series. The WDC-SILSO team has been working on this problem through the VAL-U-SUN (\url{http://www.sidc.be/valusun/}) Belgian BRAIN project, the organization of sunspot workshops (\url{ https://ssnworkshop.fandom.com/wiki/Home}), editorial work for a Solar Physics topical issue on SN re-calibration \citep{article} and more recently an ISSI team International Space Science Institute (ISSI) on recalibration of the SN (\url{https://www.issibern.ch/teams/sunspotnoser/}). \\

As the purpose of this work is to evaluate different reconstructions of a particular series of observations from the 19th century, we are not going to go beyond the notion of k-factors. 

\section{Data Sources}\label{sec:data sources}

In this section we describe in more details the sources of all the data we will be using throughout the article, by chronological order of the start of observations.\\

\subsection{Observations by Samuel Heinrich Schwabe (1825-1867)}
\label{s-Schwabe}
Samuel Heinrich Schwabe made drawings of his sunspot observations from  November 5$^{th}$ 1825 to December 29$^{th}$ 1867 covering four solar cycles: they are preserved in the library of the Royal Astronomical Society in London \citep{2011AN....332..805A}. For this observer, there are three sources of data: (1) the original observations digitized by \cite{2013MNRAS.433.3165A}, (2) the Mittheilungen \citep{1850MiZur...1....3W} data which are inconsistently marked as Schwabe's and (3) the data located in the Source Books (cf. section 3.2.2) that is explicitly marked "Schwabe".

A reconstruction of these sunspot data from Schwabe's original drawings was carried out by (\citealp{2013MNRAS.433.3165A}, A2013 henceforth). Schwabe's observation period completely overlaps with Carrington's observational years (1853-1861).

Note that in this study we only use dataset (1), i.e. Schwabe's original observations as reported in A2013 and Carrington data from the Mittheilungen as comparison. The data which appears in the Mittheilungen as Schwabe's raw data shows an inhomogeneity before and after 1849. After 1849, Wolf received data from Schwabe only on the days when Wolf missed his own observations which led to the application of a debatable scaling factor \citep{2016SoPh..291.2505F}. Hence, the series suffers from a scale discrepancy which itself requires detailed study, and as such is not suitable as a reference for comparison studies. 

\subsection{Observations by Rudolf Wolf (1849-1893)}
\subsubsection{Mittheilungen}
Rudolf Wolf observed between 1848 and 1893 but started reporting his data starting from 1849. He used Schwabe data predating his as an anchor for the start of the series now known as the international sunspot number (\url{http://www.sidc.be/silso/datafiles}). Over the years, he used several telescopes of which we were able to find clues in the Mittheilungen: 
\begin{itemize}
    \item There are observations made with the {\bf "Standard Method" or SM} in the Mittheilungen for which the instrument is referred to as the {\bf "4-foot refractor"} = \textit{` der entweder von mir oder von Herrn Meyer nach ganz entsprechender Art mit Vergrosserung 64 meines Vierfussers erhaltenen Normalbeobachtungen'} $\stackrel{\text{english}}{\longrightarrow}$ \textit{normal observations received either from me or from Mr. Meyer in quite appropriate way with enlargement $\times 64$ of my four-footer}.
    \item There are observations made with the {\bf "Portable Method" or PM}, for which the instrument is referred to as the {\bf "Parisian"} = \textit{` $2\frac{1}{2}$ füssigen Pariser-Fernrohr bei Vergrösserung 20 gemacht'} $\stackrel{\text{english}}{\longrightarrow}$ \textit{the $2\frac{1}{2}$ foot Parisian telescope with enlargement factor $\times 20$}. Also some pieces of text most probably refer to the Parisian, and indicate that {\bf k-factor(SM/PM)=1.5}: \textit{` Ein beigesetztes * bezeichnet Beobachtungen, welche ich mit dem kleinern Instrument machte, und mit 3/2 in Rechnung brachte '} $\stackrel{\text{english}}{\longrightarrow}$ \textit{A * indicates observations, which I made with the smaller instrument, and charged with 3/2.} This telescope is also referred to as a "small pocket telescope": \textit{`wenigen mit * bezeichneten Beobachtungen wurden auf Ausflugen mit einem kleinen Taschenfernrohr erhalten'}.  $\stackrel{\text{english}}{\longrightarrow}$ \textit{observations marked with * were obtained on trips with a small pocket telescope}: \textit{`wenigen mit * bezeichneten Beobachtungen wurden auf Ausflugen mit einem kleinen Taschenfernrohr erhalten'} $\stackrel{\text{english}}{\longrightarrow}$ \textit{observations marked with * were obtained on trips with a small pocket telescope}. It seems another reference "\textit{"Die mit * bezeichneten Beobachtungen sind auf Ausflugen mit einem kleinen Taschenfernrohr angestellt, und werden mittelset des Factors $\frac{3}{2}$ den übrigen homogen gemacht"}(observations marked * are made on excursions with a pocket telescope, the observations are homogenised by means of a $\frac{3}{2}$ adjustment factor) is made to this "pocket telescope" is made in the Mittheilungen.
    \item There is also reference to a similar aperture telescope, but that would be neither SM nor PM: $2\frac{1}{2}$ foot = \textit{` einem 2 1/2 Fusser bei Vergrosserung $\times 42$ gemacht'} $\stackrel{\text{english}}{\longrightarrow}$ \textit{a 2 1/2 ft made with enlargement $\times 42$}
\end{itemize}

Thus we know that R. Wolf observed with several telescopes, but we also find in the observation tables from the Mittheilungen that he used external observers (observations by colleagues from Europe mostly) and started using assistants when he introduced the portable or pocket telescope, starting in 1860-1861. The main point being that during the observation period by Carrington, Wolf was mostly observing with his main telescope (SM), while the smaller telescope (PM) was only introduced in 1860-1861 and used only sparsely until 1863. That means that our main comparison with Carrington data are the observations with the standard telescope. During 1853-1861 Wolf observed on 1927 days with his standard telescope. He started observing with his portable telescope on January 3rd 1861 and Carrington's last day of observation was on March 9th 1861: during this period Wolf made 23 observations with his portable telescope out of which 13 days overlap with Carrington's days of observations (cf. section 6.2.2).
\subsubsection{Wolf's Source Books}
\label{s-Wolf}\label{sec:wolf source books}
Wolf first recorded his observations in "Source Books" and only later sent them for typewriting in digitized format, i.e. for publication in the Mittheilungen. They are  a collection of Wolf’s own handwritten records on loose (unbinded) pages that were recovered at the ETH Library in Zurich in 2015. In fact some of the data Wolf recorded in his Source Books he did not use, and thus might not have had them printed in the Mittheilungen. This applies to a few spots recorded by Wolf for Carrington. The Source Books from 1849 to 1877 were digitized by Thomas Friedli in 2017, and appear on the Wolf Society website (\url{http://www.wolfinstitute.ch/data-tables.html}). There is apparently also data from before 1849 that has not yet been digitized. \\

The digitized Source Books allow us to : \begin{itemize}
\item Check the consistency of the data published in the Mittheilungen journals.
\item Give us the days when the data published is not from observations by Wolf himself and the names of all his auxiliary observers from 1849 onward. 
\item Find the normalization factors (k-factors) he used to scale others' data to his own observations.
\end{itemize}

\subsection{Observations by Richard Carrington (1853-1861)}

We present the different sources of data available. (Excluding the re-count by Teague of Carrington's drawings \citep{CarringtonBook1863}, see section \ref{S-aug}.

\subsubsection{Digitized catalogue of \cite{CarringtonBook1863}}
\cite{2014SoPh..289...79C}, called CV2014 henceforth, digitized the published catalogue of sunspots positions from \cite{CarringtonBook1863} in its original form. He determined the positions of 4900 sunspots (\citealp{CarringtonBook1863}, section II) between 1853 and 1861 (1853-11-17 - 1861-03-09). Since the observations were made near London, CV2014 considered the times included
in his observations to be in UT. They recomputed the heliographic longitude and latitude for each observation from the distance to the solar disc centre and the position angle and examined all observations for which there was a significant difference to correct digitization or catalogue mistakes. They found a systematic difference in longitude of about 8° (7°.99) between their recalculations and Carrington’s original data.

 When needed we have applied a correction for this error. Section \ref{sec:the longitude discrepancy} gives a plausible explanation for this longitude difference. 
\subsubsection{Carrington's data in the Mittheilungen}\label{sec:Carrington data in mitt}

In the Mittheilungen (\citealp{1865MiZur...2..193W,1874MiZur...4..173W}), Rudolf Wolf primarily published the daily group counts from Carrington's observation along with daily spot areas instead of spot counts (Mittheilungen XXXV, Rubric-303, Page- 241, published in 1874). However, for the years 1859-1860 he also published the daily spot counts along with spot areas and group counts \citep{1865MiZur...2..193W} (Mittheilungen XVII, Rubrics-199, Page-224 , published in 1865). 

The data associated to Richard Carrington in the Mittheilungen is as follows:
\begin{itemize}
    \item Group counts : 1853-11-09 $\longrightarrow$ 1860-12-26
    \item Sunspot areas: 1854-01-08 $\longrightarrow$ 1860-12-26
    \item Sunspot counts: 1859-01-02 $\longrightarrow$ 1860-12-26
\end{itemize}

Note that Carrington's data in Mittheilungen is only till 1860 even though Carrington continued his observations till March-1861. However, the Mittheilungen tables for Wolf's observations do not identify auxiliary observers who used to compute the daily sunspot number between the start of the observations in 1849 and 1859 : the first distinctions appear as tags in the annual tables in January 1859. Which means that, according to R. Wolf, there are data for R. Carrington before 1859 being used in the daily spot counts, but the Mittheilungen do not identify them. However, the Source Books mentioned in \ref{sec:wolf source books} enable the identification of 85 measurements by R. Carrington before 1859, some of them even contain a number of spots instead of sunspot area. These numbers have probably been computed from the relationship mentioned by R. Wolf in Mittheilungen XXXV as they correspond exactly to $24(Spot Area/1000) * $ (see section 5).

\subsubsection{Carrington catalogue of spots and groups by \cite{2012Ge&Ae..52..843L}}
\label{s-lepshokov}

\cite{2012Ge&Ae..52..843L}, LP2012 from now on, focus on the reconstruction of sunspot group characteristics from daily drawings and synoptic maps. We wish to compare their results to the original observations by Carrington (CV2014), by making use of the simple k-factors described above: we thus fit a slope passing through zero. For accuracy, we compute this slope using three methods: (1) the simple ordinary least square OLS(x,y), (2) the inverted OLS(y,x) and (3) a total least square fit (taking into account that neither LP2012 nor CV2014 can be considered as a reference) TLS(x,y). In all similar figures throughout the article, the slope will be determined by a weighted mean of (1), (2) and (3) with an accompanying weighted error. 

\begin{figure}[hbt!]
\vspace{10pt}
\hspace{-35pt}
\begin{minipage}{0.60\textwidth}
\centering
\includegraphics[width=6cm]{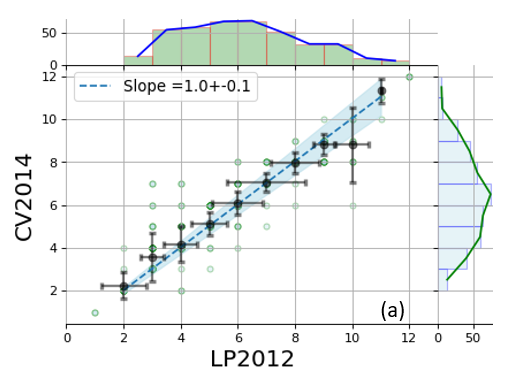}

\end{minipage}
\hspace{-40pt}
\begin{minipage}{0.60\textwidth}
\centering
\includegraphics[width=6cm]{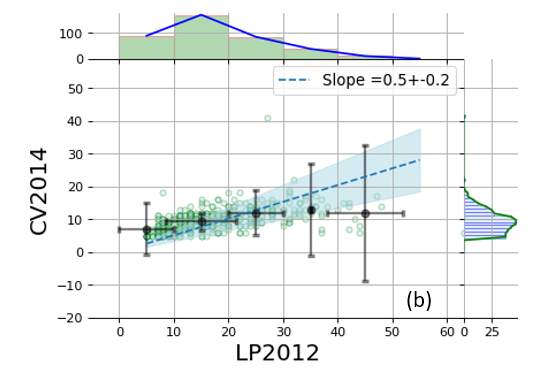}

\end{minipage}
\vspace{6pt}
\caption{Comparison of the counts for each day by Carrington as  published by LP2012 (x-axis) and CV2014 (y-axis). This figure has two panels (a for groups) and (b for spots), and each panel is composed of a central, upper and right panel. The central panel compares daily counts from the two sources as green circles with a blue regression line with error as shaded region and error bars. The upper panel gives the distribution of counts of the data from the x-axis while the right panel gives the same for the data mentioned in the y-axis. The bin size is chosen as 1 for groups and 10 for spots.}
\label{f-2}
\end{figure}

LP2012 report a similar number of groups (figure \ref{f-2}a) as Carrington's original catalogue data from CV2014, however they report larger spot counts (figure \ref{f-2}b). This difference between a modern re-count and Carrington's original counts can be attributed to the fact that Carrington was mostly interested in determining the rotation of the Sun, and chose the sunspots for an ideal placement of the “crosswire” to keep track of the groups. During minima he had few choices of spots, contrary to maximum activity. Based on our analysis of the catalogue from CV2014, Carrington had on average 2 spots per group, whereas Wolf (cf. section 3.2) and Teague (cf. section 4) had 4 spots per group - modern day counting gives about 6-7 spots per group depending on the cycle. 

\begin{figure}[hbt!] 

\centering
\includegraphics[width=10cm]{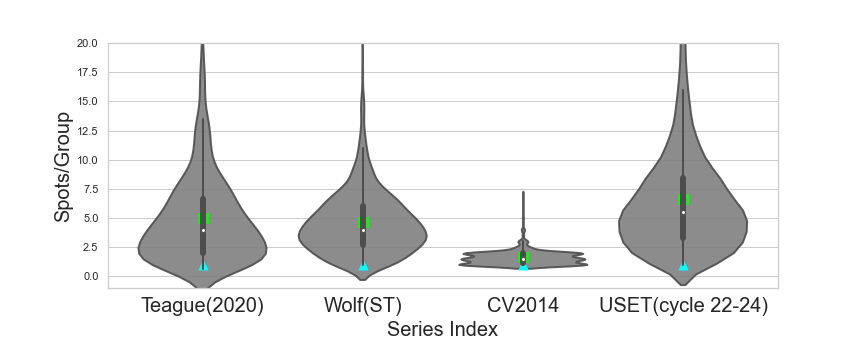}
\caption{Violin plots showing number of spots per groups for different observers. The white dot, the lime green square and cyan triangle in the center of the violin locates the median, mean and mode of the distribution respectively. The thick gray bar shows the interquartile range, and the thin gray bar depicts the interdecile range \citep{2019ApJ...886....7M}. The bin width is computed with Scott's rule \citep{2010}}.
\label{f-3}
\end{figure}

Figure \ref{f-3} presents violin plots \citep{Hintze.Nelson1998} of the number of spots per groups for the re-count by Teague (this work), the data from Wolf (SM), the original count by Carrington (CV2014) and the Uccle Solar Equitorial Table (USET) data. The median, mean and mode values are 4,5,1 for Teague (2020) respectively, 4,5,1 for Wolf (SM), 2,2,1 for CV2014 and 6,7,1 for USET for cycle 22 to 24 (September 1986 - December,2019). All values are rounded up to their nearest integer as it gives a picture of the physical meaning of the ratio "number of spots per group". This value of 2 spots per group is linked directly to the fact that for tracking a group during a rotation, Carrington chose 2 spots on average from the leading and trailing part of a group. Hence, the underestimation. In addition to that, Carrington may also have been counting each penumbra as one spot.

\begin{figure}[hbt!]
\vspace{10pt}
\hspace{-40pt}
\begin{minipage}{0.60\textwidth}
\centering
\includegraphics[width=6cm]{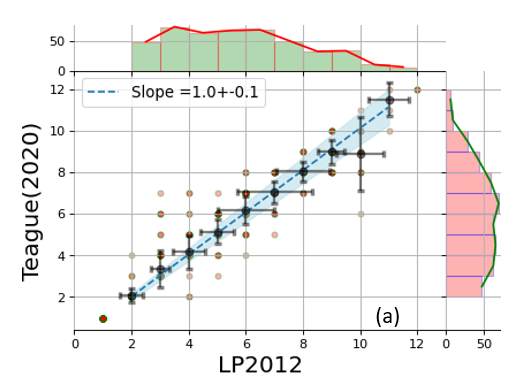}

\end{minipage}
\hspace{-40pt}
\begin{minipage}{0.60\textwidth}
\centering
\includegraphics[width=5.5cm]{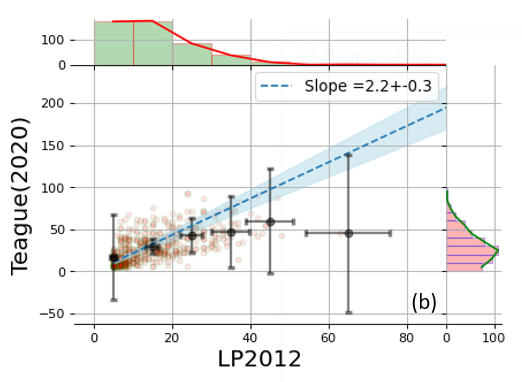}

\end{minipage}
\caption{Same as Fig. \ref{f-2} for daily counts by Carrington as  published by LP2012 (x-axis) and re-counted by Teague (2020), on the y-axis for groups (a) and spots (b).}
\label{f-4}
\end{figure}

On further comparison of the reconstruction of Carrington's daily observations by LP2012 with the reconstruction by Teague, the latter reported more spot counts compared to the former (figure \ref{f-4}a), while again, the group counts do not vary much (figure \ref{f-4}b).
The difference can be attributed to the fact that LP2012 implemented an automated detection of the drawings after digitizing the original catalog in gray scale gradation, which may have led to overlooking of a few spots during the rising phase of solar activity.

\begin{figure}

\centering
\includegraphics[width=10cm]{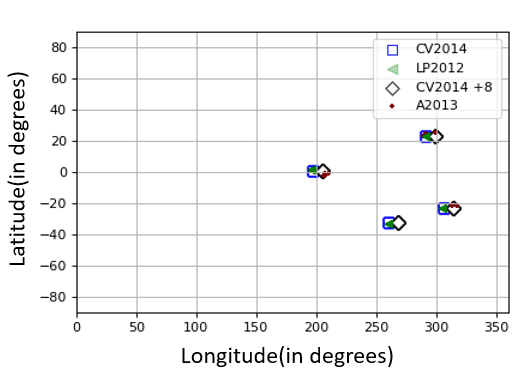}
\
\caption{ Comparison of the positions from CV2014 , LP2014 and A2013 for May 05, 1857 with a correction of +8 degrees applied to all longitudes of CV2014 as explained in their work.}
\label{f-5}
\end{figure}
In figure \ref{f-5}, we compare the positions of the daily spots/groups as reported by Carrington in his original catalog (CV2014) to the positions of Carrington's daily observations published by LP2012, and to the positions reported by Samuel Heinrich Schwabe (cf. section 3.1.) reconstructed by A2013. All positions from CV2014 and LP2012 had to be corrected by the 8 degrees mentioned in section 3.3.1. to match the data from A2013.\\

Considering this dataset suffers from the original bias in longitude and counts fewer spots than the current re-count by Teague, we are not going to use it further in this analysis which focuses on the North-South distribution of spots and groups.
\subsubsection{ The longitude discrepancy}\label{sec:the longitude discrepancy}
As pointed out in sections 3.3.1 and 3.3.3, CV2014 and LP2012 show an $\approx$8$\degree$ discrepancy in Carrington's spots longitudes compared to modern recomputations or other overlapping observers (cf. fig 5).
We found that this shift of all longitudes can be explained by two discrepancies in the base conventions:
\begin{itemize}
    \item 	There is a 12-hour difference between the epoch of coincidence of the prime meridian with the ascending node of the Sun’s equator, E, in current use (1854 Jan 1 at 1200 UT) and the epoch selected by Carrington (1854 Jan 1 at 0000 UT).
    \item Although the value for the longitude of the ascending node, N, that Carrington proposed “for future adoption” (73° 40´ for 1850.0) is in close agreement with the currently accepted value, it is \textit{not} this value that he used in calculating the sunspot positions listed in his catalogue, for which he provisionally adopted N = 74° 30' for 1854.0 (implying N $\approx$ 74° 27' for 1850.0) 

\end{itemize}

Two different conventions add up to explain the shift in longitudes:
\begin{itemize}
    \item 	With a sidereal rotation rate of 25.38 days, the 12-hour difference in the origin of times (point i) is equal to
   		14,1844° / 2 = 7.0922°  or 7° 05' 32"
   		
   	\item The difference in the longitude of the node (point ii) gives:
   		74° 30' - 73° 40'  = 0° 50' or 0.8333°
\end{itemize}

Using a full ephemeris according to \cite{10.5555/583510}, we could verify that this simple addition of two components closely matches the value of the total discrepancy (adding up those two shifts, we get 7° 55' 32" or 7,9255°.). Compared to the 7.99° value from CV2014, this leaves a small mismatch of only 3.9'.
If this small angle is translated into a linear error, for 303mm or $\approx 12$ inch diameter for the drawing Carrington used (\citealp{CarringtonBook1863}, p.8), it corresponds to 0.7mm, i.e. 0.23\% of the solar diameter (maximum at disc centre). This can be considered as a very good precision when measuring the centre of spots or groups, given their irregular shape and patterns . So, this small residual difference can indeed be considered as insignificant, even with the best precision achievable in such observations.

It is interesting to note that the projected image of the Sun Carrington used \citep{2012SoPh..280....1C} was 11 inches in diameter but probably Carrington traced the solar image on to a blank disc attached to the projection screen (\citealp{CarringtonBook1863}, p.9). Although this method is often described in amateur texts, experienced observers know that it is seldom practicable, even with a heavy, solidly mounted 19th century refractor of the kind Carrington used. Instead, he most probably projected the solar image on to the coated glass screen and made his drawing on a separate sheet of paper. He probably used the rectangular coordinates of the spots he had chosen to measure to position them correctly on the drawing, adding the other details by eye. That may seem a laborious process, but it is probably more accurate than attempting to trace the projected image directly.
\section{Modern reconstruction of (hemispheric) sunspot numbers from Carrington’s observations}\label{sec:modern reconstruction}
\subsection{Method}
By reference to Carrington’s published drawings and measurements, Teague has estimated the sunspot number according to the formula $10g + f$ for each hemisphere on each day of observation, providing a separate count for each individual group. Subject to necessary omissions (such as the quality of the ‘seeing’, which Carrington did not bother to record) he applied the standard procedure as currently used by members of the WDC-SILSO observing network.\\
Wherever available, Teague relied upon Carrington’s daily drawing sequences. For groups observed only once, he used the relevant rotation drawings. In the very rare cases where no drawing of either kind was available (usually groups extremely close to the solar limb), he referred to Carrington’s measured positions together with any additional notes.

\begin{figure}[hbt!] 

\centering
\includegraphics[width=7cm]{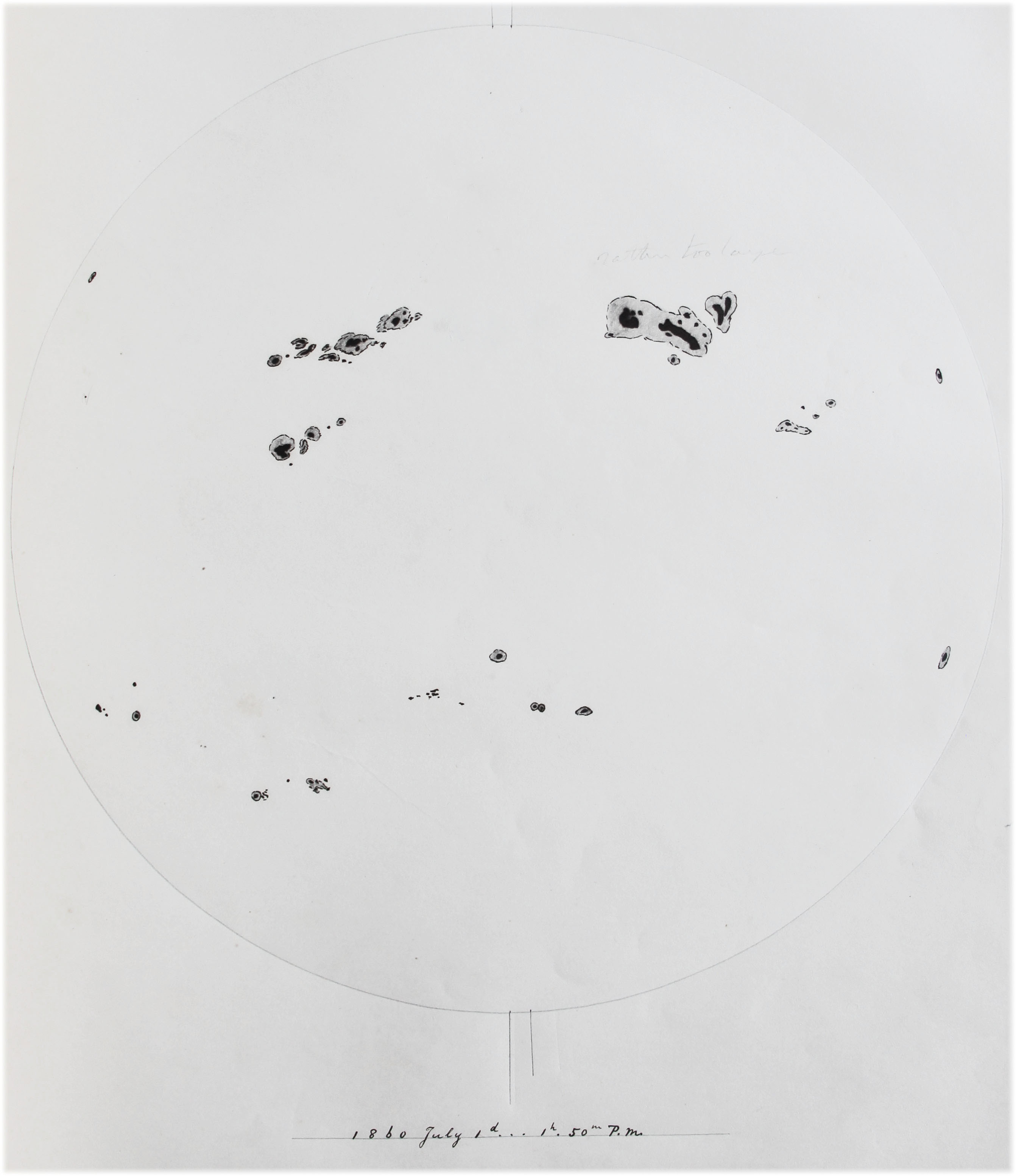}
\caption{ RAS MSS Carrington 1860-07-01: Whole disc drawing from July 1st 1860: CV2014 reports 10 groups and 22 spots while Teague (this work) reports 11 groups and 82 spots - cf. table \ref{tab:Teague_con}.}
\label{f-6}
\end{figure}

Since Carrington’s original drawings and observations remain in the hands of the Royal Astronomical Society in London, and have never been published, it would not have been practical to use them as a basis for deriving hemispheric sunspot numbers. Teague therefore relied exclusively upon the published drawings. However, a ‘dip sample’ comparison of the two sources was done, and suggests that Carrington was characteristically meticulous in transferring details from his whole disc drawings to the published figures. Figure \ref{f-6} presents an original whole-disc drawing from July 1st 1860. 

Although the counting of spots and groups on individual cases reveals occasional disparities, the overall agreement is good. In practice, the counting difference for spots between the \textbf{published drawings (P)} and in the corresponding \textbf{whole-disc drawings (WD)} does not exceed 8\% for a single day,  while the mean difference on the whole sample is less than 2\%. i.e.
$$\frac{|N_s^j(WD) - N_s^j(P)|}{\frac{1}{2n}\sum_{i=1}^n \left(N_s^i(WD) + N_s^i(P)\right)} = \frac{|N_s(WD) - N_s(P)|}{\text{mean}\left(\{N_s(WD)\} \cup \{N_s(P)\right)\}} \begin{cases} \text{max} = 8\%\quad\text{on Dec. 1st 1860} \\ \text{mean} = 2\% \end{cases}$$

From the modern researcher’s point of view, the principal value of the whole disc drawings is their contemporaneity. The principal advantages of the published sequences of drawings is their consistency (all were prepared to a uniform scale by Carrington himself), together with the fact that they help to resolve ambiguities in the original drawings (for example, whether a stippled effect is intended to depict an area of penumbra as opposed to a collection of tiny spots in close proximity to one another).

\subsection{Results: Sunspot Number for the total, north and south extracted from the data.}

  Table \ref{tab:Teague_ori} presents the data re-counted by Teague. Column 1 gives the date as recorded by Carrington, column 2 the serial number of the group assigned by Carrington. By reference to Carrington’s position measurements, column 3 specifies whether the group was in the northern (‘N’) or southern (‘S’) hemisphere.
  
   \begin{table}[H]
\addtolength{\tabcolsep}{-4.5pt}
\caption{Excerpt of the original data as re-counted by E.T.H. Teague}\label{tab:Teague_ori}

\begin{tabular}{|l|l|l|l|l|l|l|l|}
\hline
\textbf{Date} & \textbf{No} & \textbf{N / S} & \textbf{N} & \textbf{S} & \textbf{C / R} & \textbf{Wolf} & \textbf{Notes} \\ \hline
22 Apr 1860 & 693 & \textbf{S} & 0 & 12 &  &  &  \\ \hline
22 Apr 1860 & 697 & N & 14 & 0 &  &  &  \\ \hline
22 Apr 1860 & 700 & N & 11 & 0 & 37 & 60 &  \\ \hline
25 Apr 1860 & 696 & S & 0 & 12 &  &  &  \\ \hline
25 Apr 1860 & 697 & N & 13 & 0 &  &  &  \\ \hline
25 Apr 1860 & 699 & N & 12 & 0 &  &  & No sequence One measure, but rotation chart shows 2 spots \\ \hline
25 Apr 1860 & 700 & N & 13 & 0 &  &  &  \\ \hline
25 Apr 1860 & 701/4 & S & 0 & 32 & 82 & 76 & Two groups \\ \hline
28 Apr 1860 & 701/4 & S & 0 & 37 &  &  & See note for 25 April \\ \hline
28 Apr 1860 & 707 & N & 17 & 0 & 54 & 52 &  \\ \hline
29 Apr 1860 & 701/4 & S & 0 & 32 &  &  & See note for 25 April \\ \hline
29 Apr 1860 & 703 & N & 12 & 0 &  &  &  \\ \hline
29 Apr 1860 & 705 & N & 12 & 0 &  &  &  \\ \hline
29 Apr 1860 & 707 & N & 14 & 0 & 70 & 73 &  \\ \hline
\end{tabular}%

\end{table}
 
  Columns 4 and 5 list Teague’s estimated counts (10g+f) for northern and southern hemispheres. On the last line of each day’s record, column 6 gives the total sunspot count for the day, and column 7 (‘Wolf’) gives the equivalent total R count for that day taken from Waldmeier \citep{1962JRASC..56..235W}.
 Column 8 contains Teague’s comments (including selected extracts from Carrington’s own notes). Also in column 8, at the end of each month’s entries, there appears a summary in bold type, listing the mean daily R counts for the northern and southern hemispheres, the total mean daily R count (i.e. the sum of both hemispheres), and finally, for comparison purposes, the total mean daily Wolf figure as calculated from Waldmeier’s data by reference only to the particular days on which Carrington observed during the month in question (i.e. ignoring all other days in the month).
A complete version of this table is published online (\url{http://www.sidc.be/silso/carrington}).

Table \ref{tab:Teague_con} is the consolidated version of Table \ref{tab:Teague_ori}. Column 1 gives the date of observation, columns 2, 3, 5 and 6 give the groups and sunspots counts in the northern hemisphere and in the southern hemisphere respectively, on the date given in column 1. Columns 4 and 7 give Teague's estimated counts (10g+s) for the northern and southern hemispheres respectively. 

\begin{table}[H]
\addtolength{\tabcolsep}{-2.5pt}
\caption{Excerpt of the consolidated version of Table \ref{tab:Teague_ori} with shaded areas corresponding to the dates mentioned in Table \ref{tab:Teague_ori}. The black line with text in white corresponds to the date for Carrington's original drawing in figure \ref{f-6}. The dotted line represents continuation.}\label{tab:Teague_con}
\begin{tabular}{|l|l|l|l|l|l|l|l|l|l|l|l|l|}
\hline
\textbf{Date} & \textbf{G\_N} & \textbf{S\_N} & \textbf{SN\_N} & \textbf{G\_S} & \textbf{S\_S} & \textbf{SN\_S} & \textbf{G\_Tot} & \textbf{S\_Tot} & \textbf{SN\_Tot} & \textbf{SN\_V1} & \textbf{SN\_V2} & \textbf{flag} \\ \hline
1860-04-18 & 3 & 14 & 44 & 3 & 17 & 47 & 6 & 31 & 91 & 97 & 184 & 2 \\ \hline
1860-04-21 & 1 & 6 & 16 & 5 & 16 & 66 & 6 & 22 & 82 & 45 & 85 & 1 \\ \hline
\rowcolor[HTML]{C0C0C0} 
1860-04-22 & 2 & 5 & 25 & 1 & 2 & 12 & 3 & 7 & 37 & 60 & 114 & 1 \\ \hline
\rowcolor[HTML]{C0C0C0} 
1860-04-25 & 3 & 8 & 38 & 2 & 24 & 44 & 5 & 32 & 82 & 76 & 144 & 2 \\ \hline
\rowcolor[HTML]{C0C0C0} 
1860-04-28 & 1 & 7 & 17 & 1 & 27 & 37 & 2 & 34 & 54 & 52 & 99 & 2 \\ \hline
\rowcolor[HTML]{C0C0C0} 
1860-04-29 & 3 & 8 & 38 & 1 & 22 & 32 & 4 & 30 & 70 & 73 & 139 & 2 \\ \hline
1860-04-30 & 3 & 9 & 39 & 1 & 23 & 33 & 4 & 32 & 72 & 69 & 131 & 2 \\ \hline
... & ... & ... & ... & ... & ... & ... & ... & ... & ... & ... & ... & ... \\ \hline
1860-06-26 & 5 & 31 & 81 & 4 & 34 & 74 & 9 & 65 & 155 & 164 & 312 & 1 \\ \hline
\rowcolor[HTML]{000000} 
{\color[HTML]{FFFFFF} 1860-07-01} & {\color[HTML]{FFFFFF} 7} & {\color[HTML]{FFFFFF} 51} & {\color[HTML]{FFFFFF} 121} & {\color[HTML]{FFFFFF} 4} & {\color[HTML]{FFFFFF} 31} & {\color[HTML]{FFFFFF} 71} & {\color[HTML]{FFFFFF} 11} & {\color[HTML]{FFFFFF} 82} & {\color[HTML]{FFFFFF} 192} & {\color[HTML]{FFFFFF} 184} & {\color[HTML]{FFFFFF} 350} & {\color[HTML]{FFFFFF} 2} \\ \hline
1860-07-03 & 5 & 59 & 109 & 3 & 11 & 41 & 8 & 70 & 150 & 165 & 314 & 1 \\ \hline
1860-07-04 & 5 & 59 & 109 & 4 & 12 & 52 & 9 & 71 & 161 & 147 & 279 & 1 \\ \hline
1860-07-06 & 5 & 43 & 93 & 4 & 21 & 61 & 9 & 64 & 154 & 114 & 217 & 1 \\ \hline
1860-07-08 & 4 & 27 & 67 & 2 & 8 & 28 & 6 & 35 & 95 & 92 & 175 & 1 \\ \hline
\end{tabular}%
\end{table}
Columns 8,9 10 are groups, sunspots and estimated counts in total for the given date. Columns 11 and 12 give the International Sunspot Number version 1 and 2 respectively (\url{http://www.sidc.be/silso/home}). Column 13 corresponds to flagged data: if Teague reports any discrepancy with the counts of groups in Carrington's drawings it is flagged 2, otherwise 1 for clean data. For example, on April 25th 1860, Carrington counted group no.701 as one group but Teague reports it as 2 groups (Table \ref{tab:Teague_ori}) thus it is flagged as 2 in table \ref{tab:Teague_con} for the same date.

\begin{figure}[H]    
\centerline{\includegraphics[width=10cm]{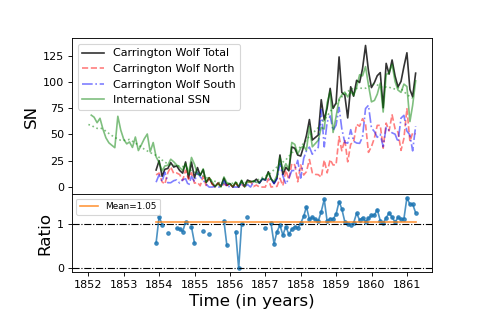}}
\caption{Wolf Number (10g+f) computed from the north and south values for Carrington re-counted by Teague (2020) with the International Sunspot Number (WDC-SILSO, ISN). The lower panel gives the ratio of Teague (2020) Total Sunspot Number / ISN.}
\label{f-7}
 \end{figure}

Teague’s original estimates for each group in both hemispheres and the consolidated table of daily estimates can be found at (\url{http://www.sidc.be/silso/carrington}). We compare the total, north and south sunspot numbers extracted from the re-count by Teague with the international sunspot number over the same period in figure \ref{f-7} .
\vspace{-3mm}

%
\begin{figure}[H]
\centerline{\includegraphics[width=15cm]{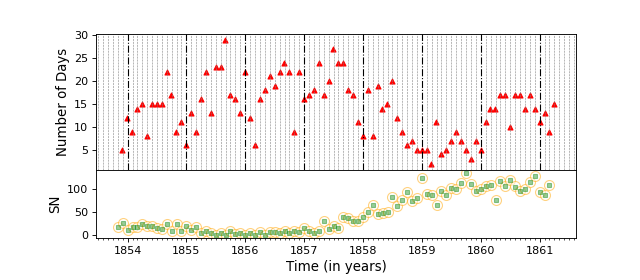}}

\caption{Number of days of observation by Carrington for each month (upper panel) and monthly Wolf numbers : orange circles represent monthly values reported by Teague and green squares represent monthly values calculated from daily values in Table 2. (lower panel). }
\label{f-8}
\end{figure}
\vspace{3mm}

 Figure \ref{f-8} shows the monthly means as computed by Teague and recomputed from the extracted daily values as well as the number of observations per month. A casual glance at figure \ref{f-8} shows the dramatic interruption to Carrington’s project during the second half of 1858, presumably caused by the personal and commercial upheaval that followed his father’s death.\\

\newpage

\section{Reconstructions by Wolf and Teague compared to Carrington's original data}
\label{s-com}
\subsection{First reconstruction by Rudolf Wolf (1859-1860)} 
\label{s-4.1}

%
%
%

\begin{figure}[hbt!]    
\begin{minipage}{0.50\textwidth}
\centering
\includegraphics[width=5cm]{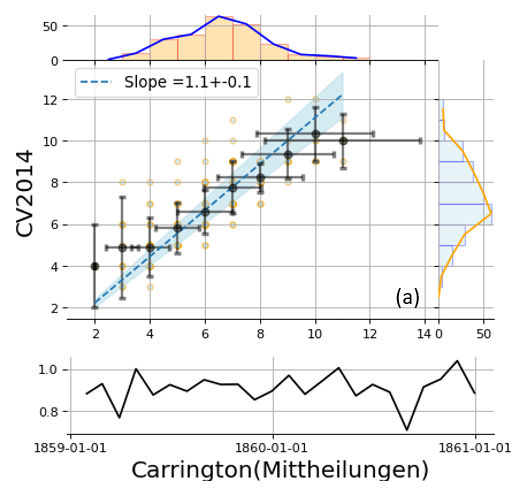}

\end{minipage}
\hspace{-10pt}
\begin{minipage}{0.50\textwidth}
\centering
\includegraphics[width=5cm]{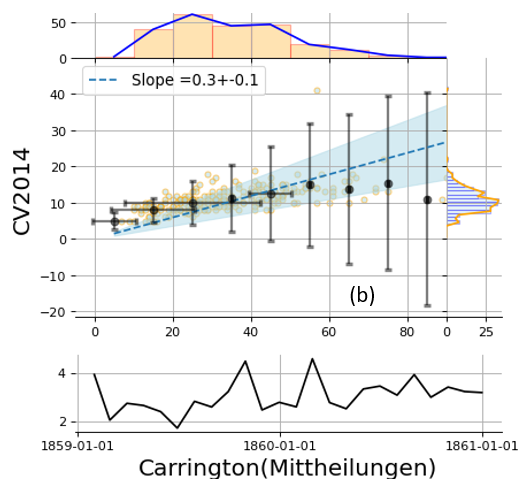}

\end{minipage}
\vspace{1pt}
\caption{Comparison of the distributions of the number of groups (a) or spots (b) reported for Carrington in CV2014 and in the Mittheilungen over 1859-1860 with the same model as Figs. 2 and 4. This is a zoom-in as spot counts go to about 100. The lower panel gives the monthly ratios based on the x-axis/y-axis. }
\label{f-9}
 \end{figure}

 Figure \ref{f-9} shows the distribution of groups and spots counts for Carrington from CV2014 versus the Mittheilungen over 1859-1860. The groups counts are similar, with a ratio of 1.1$\pm$0.1 while the spot counts in the Mittheilungen are $\approx 3$ times more numerous than in CV2014. The data from Carrington that appears in the Mittheilungen for 1859-1860 is clearly different from the original data as published in \cite{CarringtonBook1863}.

\begin{figure}[hbt!]    
\centerline{\includegraphics[width=12cm]{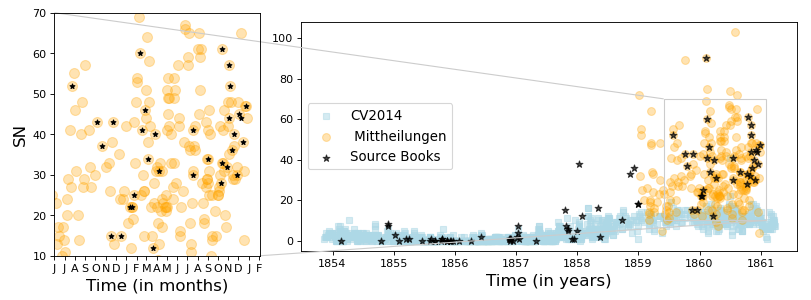}}
\caption{Comparison of the number of spots reported by Carrington in the Mittheilungen, in the original catalog CV2014 and in the Source Books for the entire period with a zoom-in.}
\label{f-10}
 \end{figure}

In  figure \ref{f-10} we compare the number of spots recorded in the Mittheilungen and in CV2014 with the number of spots identified for Carrington in Wolf's source books. The number of spots published in the Mittheilungen for 1859-1860 is consistent with the numbers from the Source books. In addition, Wolf mentions in his source books a k-factor for Carrington data of 1.03, which implies his version of Carrington's data does not deviate much from his own observations.

\begin{figure}[hbt!]    
\centerline{\includegraphics[width=12cm]{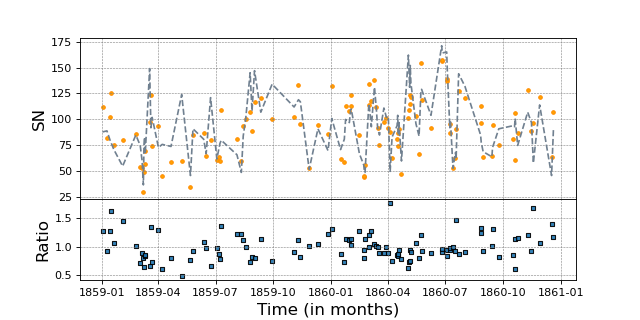}}
\caption{Upper panel: Comparison of the Wolf number (WN=10g+f) of Carrington's Mittheilungen observations (WN(Carrington), orange dots) and Wolf's own WN with the standard telescope (WN(Wolf-SM), dashed line). Lower panel:  Daily ratio between WN(Wolf-SM) WN(Carrington)on overlapping days. }
\label{f-11}
 \end{figure}

\begin{figure}[hbt!] 
\vspace{10pt}
\centering
\includegraphics[width=10cm]{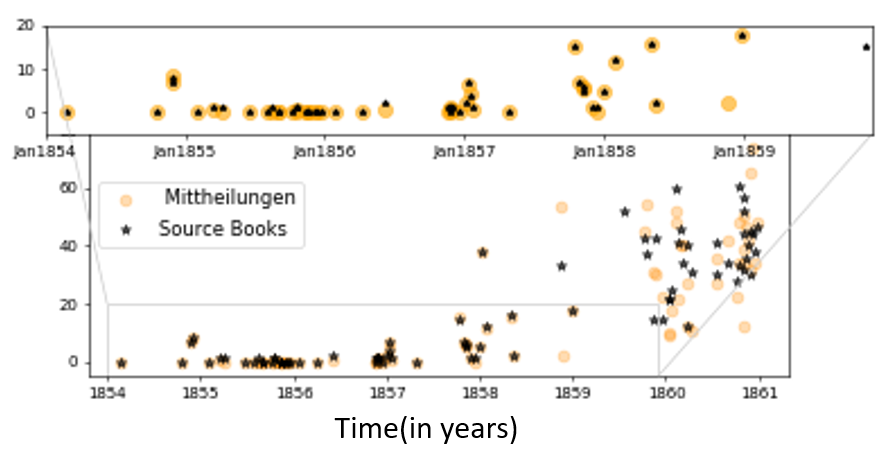}
\caption{Number of spots calculated from the spot area information of Carrington and number of spots used by Wolf as mentioned his source books from Carrington's observations, focusing on the period 1854-1859. The x-axis in the zoomed plot represents time and the y-axis number of spots. }
 \label{f-12}
\end{figure}

Figure \ref{f-11} shows the Wolf Number from the Mittheilungen for Carrington and Wolf himself over 1859-1860, as well as their ratio. It is indeed compatible with a k-factor of 1.03. There are two possible explanations for this: the observations that appear in the Mittheilungen under Carrington are (1) a re-count of the original drawings by Wolf himself or (2) a reconstruction of the spot counts from the available areas. The Mittheilungen state that the 2 years worth of group and spot counts were obtained from the book by Carrington in 1865. And additionally the Mittheilungen reveal that Wolf had derived a relationship areas to spot counts, that he most probably applied to the years before 1859 (figure \ref{f-12}).

We surmise that around 1860, Wolf visited London and came across Carrington's observations. He could access a few of his daily drawings and calculated the k-factor of 1.03. Possibly on his request later, Carrington sent his original drawings for the years 1859-1860 which were published in Mitteilungen XVII (1865) and figure \ref{f-10} shows these were the exact counts which were used to fill Wolf's observational gaps during the same period.

Later on, Wolf got access to Carrington's logbooks where he reported daily areas of the spots for the period 1853-1861 \citep{1874MiZur...4..173W}. He developed a relation that appears in Mittheilungen XXXV that relates spots areas and spot numbers as $Spot Number = 24(Spot Area/1000)$. We applied this relation to spot areas in the Mittheilungen and figure \ref{f-12} shows that Wolf actually calculated the number of spots until 1858 using the relation mentioned above and computed from the spot counts acquired in 1865. Therefore, the spot counts used by Wolf as Carrington's in his source books, were actually reconstructed spot numbers from 1853-1858 from area information and actual re-counts from drawings during 1859-1860.
\subsection{Reconstruction by Teague}  
\label{s-4.2}

\subsubsection{Numbers of groups and spots}
\label{s-4.2.1}
We compare the original counts between CV2014 and the re-count by Teague as it embodies the difference in bias (and in knowledge) between the 1850s and today, especially since Teague is an active observer of the WDC-SILSO network.
\begin{figure}[hbt!] 
\begin{minipage}{0.50\textwidth}
\centering
\includegraphics[width=5cm]{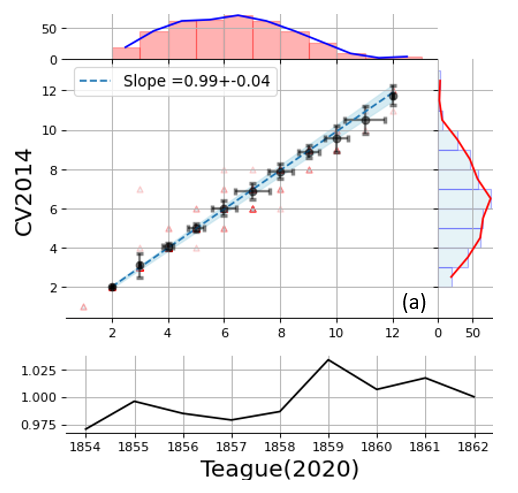}

\end{minipage}
\hspace{-80pt}
\begin{minipage}{0.50\textwidth}
\centering
\includegraphics[width=5cm]{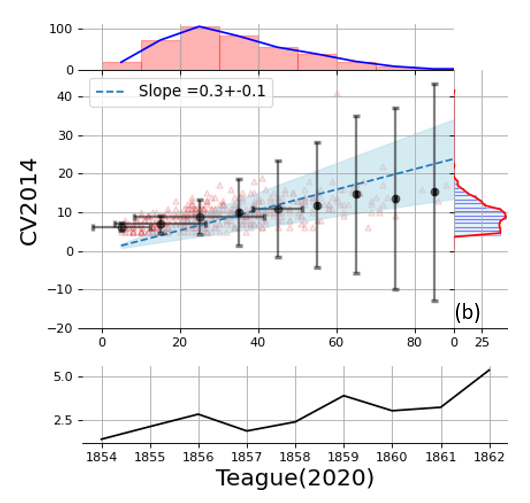}

\end{minipage}
\vspace{-7pt}
\begin{minipage}{0.99\textwidth}
\centering
\includegraphics[width=8cm]{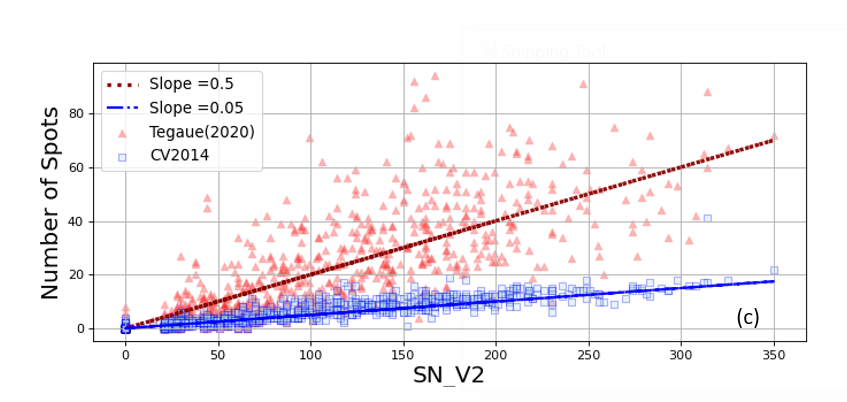}

\end{minipage}
\vspace{2pt}
\caption{Comparison of number of groups(a) and spots (b) from CV2014 vs. Teague (2020) with the same model as Fig. 9. The lower panel gives the monthly ratios based on the x-axis/y-axis. (c) shows the evolution of spot counts for both datasets with ISN(V2).}
\label{f-13}
\end{figure}

From figure \ref{f-13}a we see that the group counts are in agreement but the sunspot counts are not (cf. \ref{f-13}b). This discrepancy in spot counts is most probably due to the fact that CV2014 is a position catalog (cf. section \ref{s-lepshokov}). Carrington was mostly interested in determining the rotation of the sun, so he chose ideal sunspots for placing the “crosswire” to keep track of a group. A time based analysis of the ratio of CV2014 counts to Teague counts is shown in figure \ref{f-13}a,b in the lower panels for groups and spots counts. It shows that the scale evolves over time for spots while it stays around 1 for groups. During minima (1853-1857), Carrington had no choice but to track every sunspot due to their scarcity, hence the ratio lies close to 1 (figure \ref{f-13}b), but as the activity increases, he had  more spots to choose from and thus, the ratio largely deviates from 1. For the tracking of a group during a rotation, Carrington chose 2 spots on average from the leading and trailing part of a group (thus 2 spots per group, cf. figure \ref{f-3}). It is evident from figure \ref{f-13}c that with the increase in International Sunspot Number (proxy of solar activity) the spot counts for Teague(2020) increases as expected but spot counts for CV2014 does not follow the same trend, confirming the above hypothesis. 

\subsubsection{Hemispheric numbers of groups and spots}
\vspace{-10pt}

\begin{figure}[hbt!]
\begin{minipage}{0.50\textwidth} 
\begin{minipage}{0.99\textwidth}
\centering
\includegraphics[width=5cm]{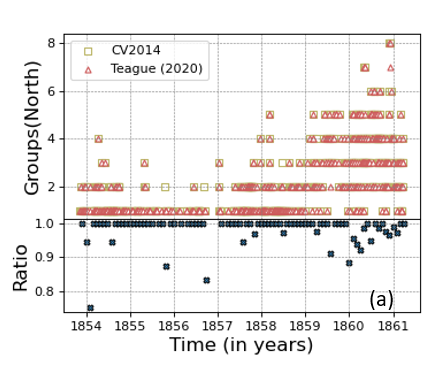}
\end{minipage}
\vspace{5pt}
\begin{minipage}{0.99\textwidth}
\centering
\includegraphics[ width=5cm]{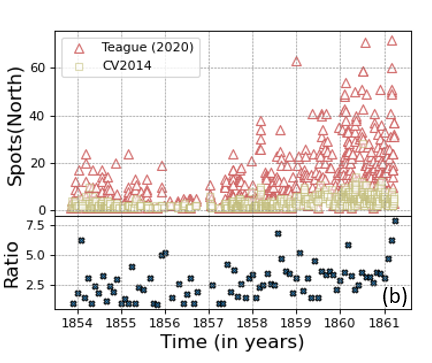}

\end{minipage}
\end{minipage}
\hspace{-10pt}
\begin{minipage}{0.50\textwidth}
\begin{minipage}{0.99\textwidth}
\centering
\includegraphics[width=5cm]{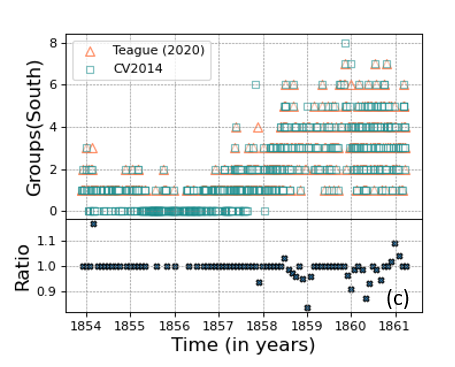}

\end{minipage}
\vspace{5pt}
\begin{minipage}{0.99\textwidth}
\centering
\includegraphics[width=5cm]{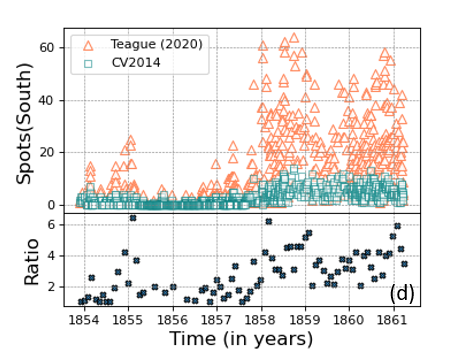}

\end{minipage}
\end{minipage}
\vspace{5pt}
\caption{Comparison of hemispheric numbers of groups and spots as counted by Carrington and published by CV2014 and re-counts of the original drawing by Teague.(a) and (b) represents the group and spot counts respectively in the Northern Hemisphere,(c) and (d) represents the the same for the Southern hemisphere. The lower panels represent monthly averages of the ratio Teague vs CV2014.}
\label{f-14}
\end{figure}
 We compare the hemispheric groups and spots numbers from the re-count of Carrington's data by Teague and Carrington's original observations (CV2014), to expose any bias that may exist at the hemispherical level.\\
 
  The group counts match almost perfectly in both hemispheres in both catalogs (figure \ref{f-14}a and c). The spot counts in CV2014 are underestimated in both hemispheres (figure \ref{f-14}b and d), similarly to the whole data as shown in figure \ref{f-13}b. However, the number of spots counted by Teague(2020) and CV2014 show much higher values in the Southern Hemisphere. This is coherent with the fact that during cycle 10 (1855 December – 1867 March) the solar activity of the Southern Hemisphere was higher by approximately 5\% in sunspot number \citep{2012Ge&Ae..52..843L}. \\

This behaviour is in line with the conclusions based on total sunspot numbers and the dominance of the southern hemisphere during this specific solar cycle.

\section{Comparison of reconstructions with observations by Wolf and Scwhabe}\label{sec:comparison obs}
From figure \ref{f-1}, it is evident that the observers who overlap Richard Carrington's entire period of observation (1853-1861) are Samuel Heinrich Schwabe (1825-1868) and Rudolf Wolf (1849-1893). We restrict our study in this article to the observers who entirely overlap with Carrington's observation period. Hence, in this section we present the comparison study of Carrington's original observations (CV2014) and the reconstruction by Teague with reconstructed observations of Samuel Heinrich Schwabe by A2013 and original observations of Pr. Rudolf Wolf (Mittheilungen and Source Books).
\subsection{Comparison with Rudolf Wolf's observation}

\subsubsection{Comparison with Rudolf Wolf's Standard Telescope observations} 
\label{s-5.2.1}

We compare the distribution of spots and groups counts of Wolf's observations with his standard telescope (cf.section-\ref{s-Wolf}) with the reconstructions of Carrington's observation by Wolf (1874) and Teague (2020). 

\begin{figure}[hbt!] 
\vspace{5pt}
\hspace{-15pt}
\begin{minipage}{0.50\textwidth}
\begin{minipage}{0.99\textwidth}
\centering
\includegraphics[width=5.5cm]{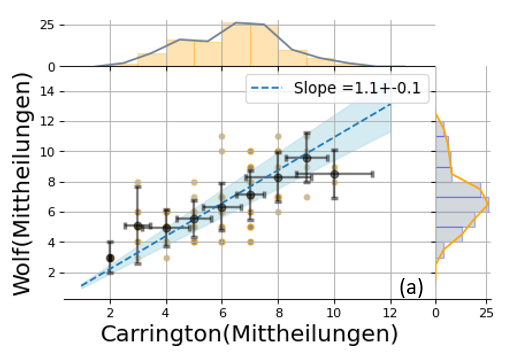}

\end{minipage}
\end{minipage}
\begin{minipage}{0.50\textwidth}
\begin{minipage}{0.99\textwidth}
\centering
\includegraphics[width=5.5cm]{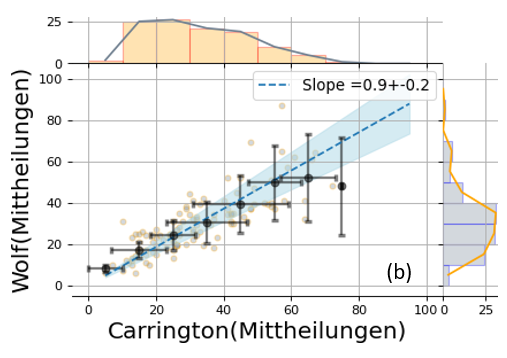}

\end{minipage}
\end{minipage}
\begin{minipage}{0.50\textwidth}
\begin{minipage}{0.9\textwidth}
\centering
\includegraphics[width=5.5cm]{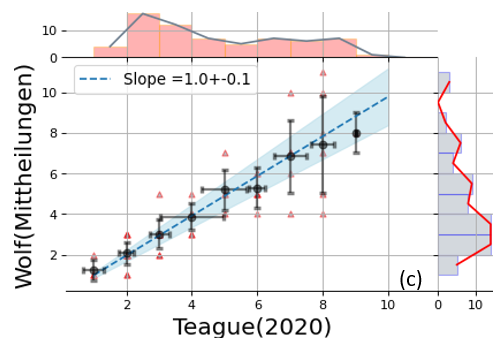}

\end{minipage}
\end{minipage}
\begin{minipage}{0.50\textwidth}
\begin{minipage}{0.9\textwidth}
\centering
\includegraphics[width=5.5cm]{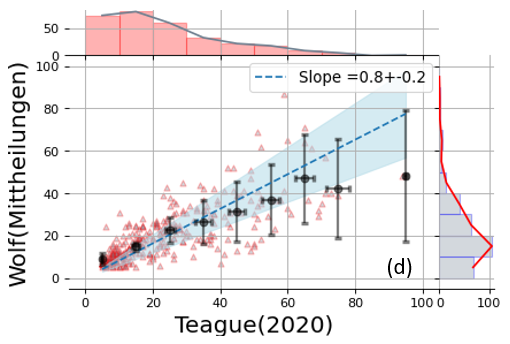}

\end{minipage}
\end{minipage}
\begin{minipage}{0.50\textwidth}
\begin{minipage}{0.99\textwidth}
\centering
\includegraphics[width=5.5cm]{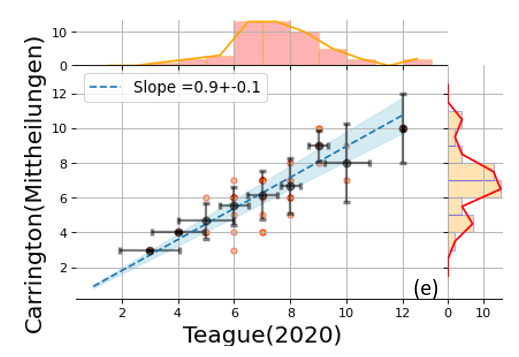}

\end{minipage}
\end{minipage}
\begin{minipage}{0.50\textwidth}
\begin{minipage}{0.99\textwidth}
\centering
\includegraphics[width=5.5cm]{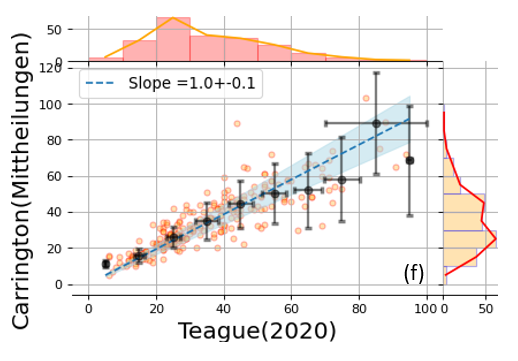}

\end{minipage}
\end{minipage}
\vspace{ -5pt}
\caption{Comparison of counts from different observers for spots and groups. All panels are presented on the same model as Fig. 15. Panels (a) and (b) present the distribution of groups and spots reported by Rudolf Wolf as observed using his standard telescope, published in the Mittheilungen and counts of Carrington published in Mittheilungen for the year 1859-1860. Panels (c) and (d) present groups and spots counts from Wolf compared to Carrington's re-counted by Teague for the period 1853-1861. Panels (e) and (f) present the distribution of the number of groups and spots published in the Mittheilungen for Carrington compared to the re-count by Teague (2020).}
\label{f-16}
\end{figure}
As established in section \ref{s-4.1} the spot counts that appears in the Mittheilungen for the years 1859 and 1860 
are re-counts of Carrington's observations by Wolf. This fact is validated by figure \ref{f-16}a and  b, where it is evident that the counts of Carrington (1859-1860) that were published in the Mittheilungen are identical (k factor consistent with 1 within error bars) to Wolf's own counts from the observations made with his standard telescope.\\

It is interesting to note that, the distribution of groups and spots counts re-counted by Teague also agree well with the distribution of Wolf's standard telescope observations (figure \ref{f-16} c and d).\\

As stated in section \ref{s-4.1}, the spot counts of Carrington that appear in \cite{1865MiZur...2..193W} for the year 1859 and 1860 (Rubrics 225, Page 199) do not correspond statistically to the original observations of Carrington, but a re-count of Carrington's sunspot observations (or a reconstruction of spot numbers from respective spot areas).
Figure \ref{f-16} e and f confirms that Teague re-counted Carrington's sunspot observations in a very similar way than Wolf when he did the first re-count of Carrington's data. Moreover, the re-counts by Teague provide data for the period when Wolf only published the spot areas along with group numbers in the Mittheilungen \citep{1874MiZur...4..173W}.
Hence, the method used by Teague for the re-counting of Carrington data gives a reliable series consistent with Wolf's counting method.

\subsubsection{Comparison with Rudolf Wolf's Portable Telescope observations}

We compare the overlapping days of observations by Wolf with his portable telescope \citep{1878MiZur...5..165W}, with Carrington's CV2014 counts and re-count by Teague. The data is not sufficient for any concrete inference of the scale discrepancy as there are only 13 effective days of overlap. Wolf applied a k-factor of 1.5 for his portable telescope to match the scale of his standard refractor. 

\begin{figure} [hbt!]  
\centerline{\includegraphics[width=12cm]{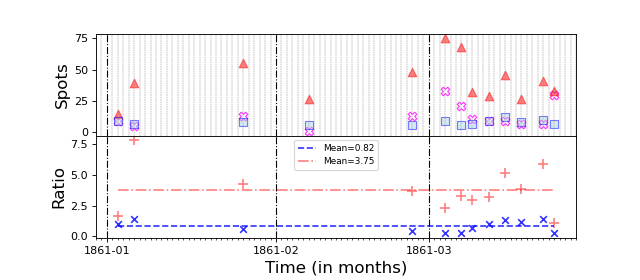}}
\caption{Upper panel: Number of spots observed by R. Wolf (Mittheilungen, PM, magenta sqaures), original counts by Carrington (CV2014, blue squares) along with spot counts re-counted by Teague (red triangles) on overlapping days. The grid lines represent each day of observations. Lower panel: red and blue crosses correspond to the ratio(i/Wolf(PM) where i= Teague(2020) or CV2014).}
\label{f-17}
\end{figure}
%
%

It is evident from figure \ref{f-17} that Teague counts more spots than Wolf with his potable telescope, as expected. It is also interesting to note that the original observations of Carrington (CV2014) almost matches with the scale of Wolf's observations with his portable telescope (figure \ref{f-17}). However, Wolf used a scaling factor of 1.5 for his portable telescope compared to 1.03 for his version of Carrington's data  (cf.section \ref{s-4.1}). This again, tends to prove the fact that Carrington's observations published in the Mittheilungen were not his original counts but a re-count by Wolf himself.

Note that in figure \ref{f-17} we omitted the data point of 1861-02-07 in the calculation of mean ratios in the lower panel for consistency purposes, as on this day, Wolf with his portable telescope reports 1 spot/1 group, CV2014 report 6 spots/4 groups whereas Teague (2020) reports 27 spots/4 groups. This discrepancy is certainly a faulty observation by R. Wolf, possibly the result of bad weather conditions.

\subsection{Comparison with Schwabe's observations}
\label{s-5.1}

\begin{figure}[hbt!]
\vspace{10pt}
\begin{minipage}{0.50\textwidth}
\begin{minipage}{0.99\textwidth}
\centering
\includegraphics[width=5.5cm]{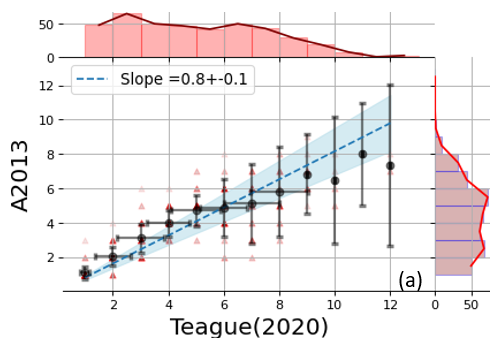}

\end{minipage}
\end{minipage}
\hspace{-5pt}
\begin{minipage}{0.50\textwidth}
\begin{minipage}{0.99\textwidth}
\centering
\includegraphics[width=5.5cm]{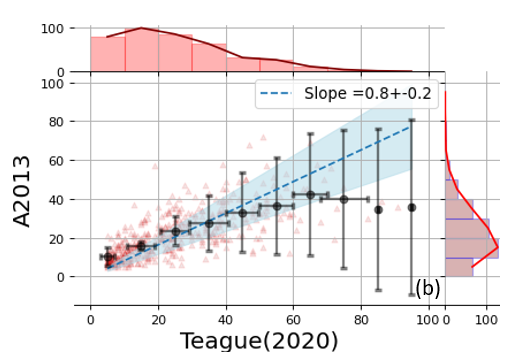}

\end{minipage}
\end{minipage}
\caption{Number of groups(a) and spots(b) from Schwabe drawings re-counted by A2013 and re-counted counts of Carrington by Teague(2020) with the same model as Figs. 2 and 4.
}
\label{f-15}
\end{figure}
We compare Schwabe's reconstructed data by A2013 (cf.section \ref{s-Schwabe}) with the reconstruction of Carrington's data by Teague (cf. Figure \ref{f-15} ).
The large error bars in the figure in bins of higher values are caused by a lack of proper statistics: the period considered (1853-1861) only has three years (1857-[March]1861) of rising solar activity and hence a small number of high values. In addition, the use by different observers of different methods of observation and different telescopes, not to mention differences in visual acuity, takes a larger importance during maximum solar activity. 

Figure \ref{f-15}, shows that the group and spots counts do not deviate much in either of the catalogs (reconstructed Carrington  catalog by Teague and reconstructed Schwabe data catalog, A2013) and also viz. original Carrington catalog (CV2014) (cf. figure 12) for the groups counts. As Teague's reconstruction is very close to Wolf's reconstruction (figure \ref{f-16}e,f), we can conclude that Carrington's reconstructed data (Wolf and Teague) basically gives the same sunspot information as Schwabe data over 1853-1861.

Therefore, the reconstruction of Carrington's data by Teague does not overestimate the spot counts and is a suitable alternative for studies involving Carrington's data in the future. 

%
%

\subsection{Scale homogeneity of the reconstructed data by Teague}  

For a series to be considered stable and homogeneous throughout its time period,it should show minimum fluctuations with respect to a stable reference series. Here, we attempt to confirm the homogeneity of the series re-counted by Teague from Carrington's original drawings, following the method developed by \cite{2019ApJ...886....7M}. 

The study by \cite{2019ApJ...886....7M} is based on the time period from 1947-2013 with a network mean of the 21 most-stable stations of the WDC-SILSO network as the reference series. We adapt the method to our dataset with a few changes, which are explained in due course. For a robust reference series, we consider the median of the observers who overlap Carrington's observation period. From figure \ref{f-1} it is evident that Rudolf Wolf and Schwabe are the only observers whose observation period overlaps the entire Carrington period. Note that we do not take into account the partial overlaps from other contemporary observers that can be seen in figure \ref{f-1}, to avoid any unnecessary biasing of the reference series for certain years. Hence, the series considered for calculation of the median series are: Wolf’s observation from Source books and Schwabe’s observation recounts by A2013. 

\begin{figure} [hbt!]  
\centerline{\includegraphics[width=12cm]{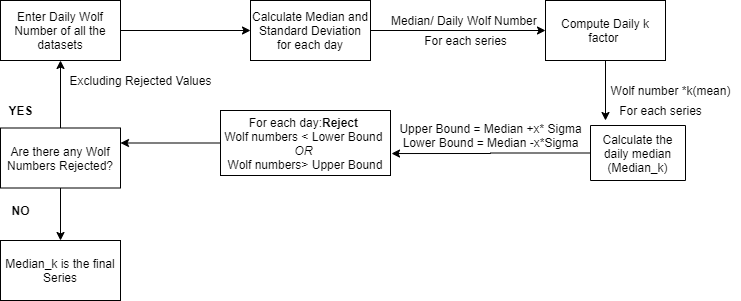}}
\caption{Flowchart showing the method for calculation of reference median series from the selected overlapping series adapted from \cite{2007AdSpR..40..919C}}
\label{f-18}
\end{figure}

\begin{figure}[hbt!] 
\vspace{5pt}
\hspace{-15pt}
\begin{minipage}{0.40\textwidth}
\begin{minipage}{0.99\textwidth}
\centering
\includegraphics[width=4cm]{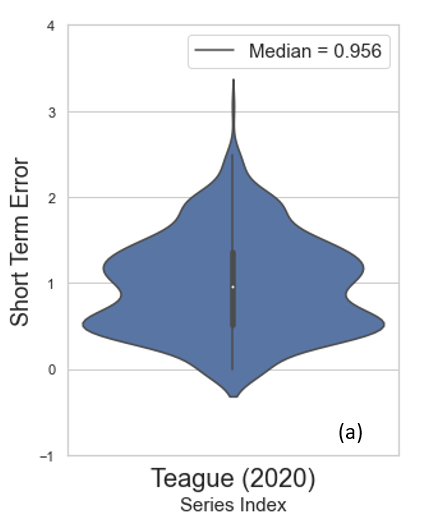}

\end{minipage}
\end{minipage}
\hspace{10pt}
\begin{minipage}{0.60\textwidth}
\begin{minipage}{0.99\textwidth}
\centering
\includegraphics[width=7cm]{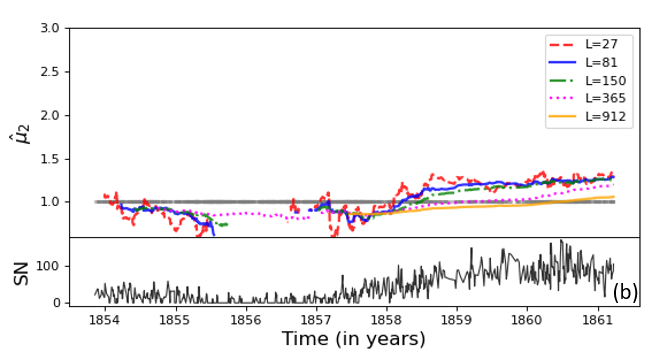}

\end{minipage}
\end{minipage}
\vspace{ 10pt}
\caption{
 (a) Violin plot of the estimated short term variability $\hat{\tilde{\epsilon}}$ for Wolf number for the series re-counted by Teague from Carrington's original drawings. The white dot in the center of the violin locates the median of the distribution. The thick gray bar shows the interquartile range, and the thin gray bar depicts the interdecile range. The bin width is computed with Scott's rule.(\cite{2019ApJ...886....7M})
 (b) Estimation of $\hat{\mu_2}$(i,t) for the series re-counted by Teague,with different MA window lengths. The solar cycle is represented in the bottom. }
\label{f-19}
\end{figure}

We adapted a simplified version of the method explained in \cite{2007AdSpR..40..919C} to calculate the reference median series as explained in figure \ref{f-18}. We chose the bound limit to be 3$\sigma$ whereas in \cite{2007AdSpR..40..919C} it is taken as 2$\sigma $, to compensate lack of availability of very stable observations.
Next,in accordance with  \cite{2019ApJ...886....7M} ,the short term error $\hat{\tilde{\epsilon}}$ is calculated by:
\begin{equation}
\label{e-1}
\hat{\tilde{\epsilon}}(i,t) = \frac{Y_i(t)}{\hat{\mu_s}(t)} 
\end{equation}
where:
\begin{equation}
\hat{\mu_s} = T(M_t)  
\end{equation}
where $Y_i(t)$ is the daily  Wolf number of the series whose short term error is to be determined, i represents the $i^{th}$ station, but here we determine the short term error for Teague series, and $\hat{\mu_s}(t)$ is the solar estimator. In \cite{2019ApJ...886....7M} $Y_i(t)$ is referred as $Z_i(t)$ as the series considered in the study require a proper scaling before estimation of the short term error, with respect to the solar estimator. In this study, we already established in section \ref{s-5.1} and section \ref{s-5.2.1} the test series (re-counts by Teague) follows a very similar distribution as the series considered for determination of our solar estimator (Observations of Wolf and Schwabe), hence, the estimation of a proper scaling factor is omitted. Therefore, we use the raw counts of Teague as $Y_i(t)$. The solar estimator is calculated by a transformation process( Network Median -> Anscombe Transformation -> Fast Fourier Transform -> Wiener Filter -> Reverse Fourier Transform-> Reverse Anscombe-> Solar Estimator) as explained in \cite{2019ApJ...886....7M}, on network median $M_t$.
Using Equation \ref{e-1} 
, the short term error for the re-counted Carrington series by Teague is calculated and represented as a violin plot in figure \ref{f-16}a. It is evident from figure \ref{f-19}a the median of the short-term error estimator $\hat{\tilde{\epsilon}}(i,t)$ is almost equal to 1 which proves the series is homogeneous with respect to the chosen solar estimator. The variance of the series is evidently on a higher side which can be attributed to the variability of the solar cycle.
For assessing the stability of the series, we use the long-term error determination from \cite{2019ApJ...886....7M}. As explained in \cite{2019ApJ...886....7M}, long term error is given by:
\begin{equation}
\label{e-2}
\hat{\mu_2}(i,t) = \left(\frac{Y_i(t)}{M_t}\right)^* 
\end{equation}
where $Y_i(t)$ are the raw daily counts from Teague, $M_t$ is the network median (as explained above) and * denoted the smoothing process involving moving average over L days. The moving average windows (MA) are chosen such that L is more than one solar rotation (27 days). We chose MA windows of lengths : 27days, 81 days, 150 days, 1 year and 2.5 year \citep{2019ApJ...886....7M}. The stability of the series is characterised by closeness of $\hat{\mu_2}$ to 1. The closer it remains to 1 , more stable is the series. It is evident from figure \ref{f-19}b, the series does not deviate from it's scale before 1858 but there is an upward trend around 1858 for the MA windows 27,81 and 150 after which again it becomes stable. This upward trend can be again, due to the rise of solar activity, as before 1858, the period was going through a minimum phase, thus, excess of 0 counts and no-observation days. Moreover, the network median we computed involves Wolf's observations, which itself suffers from a scale discrepancy around 1860-1861 due to the introduction of a lot of changes \citep{2016SoPh..291.2505F}, which can explain the little dip around the same period in figure \ref{f-19}b. Considering L>150 days, the series shows stability throughout.
\section{Conclusions}\label{sec:conclusions}
\label{s-6}
From this study, it is evident that, in his original catalogue \citep{2014SoPh..289...79C}, {\bf Richard Carrington recorded only the sunspots necessary for the positional tracking of the different groups}, because his main goal was to determine the rotation of the Sun. That is why the difference in group counting between the original method and the modern method is negligible while the spots counting sports a factor of 3 difference. \\  

Contrary to \cite{2014SoPh..289...79C} and \cite{2012Ge&Ae..52..843L}, {\bf we account for the unexplained $7.99\degree$ discrepancy in longitude between the original catalogue and the modern re-computations}. It is caused by two effects: (1) a difference between the epoch of coincidence of the prime meridian with the ascending node of the Sun’s equator used today, compared with the one selected by Carrington; (2) the value for the longitude of the ascending node proposed by Carrington is not the value he used in calculating the sunspot positions listed in his catalogue.\\

The first reconstruction of Carrington's observational data that includes sunspot numbers was carried out by Rudolf Wolf for the years 1859-1860. Wolf re-counted the sunspots and groups from the whole-disc drawings that he received directly from Carrington. From this, he extrapolated a relation between sunspot areas and number of counts, which enabled him to include infrequent Carrington numbers of spots and groups in the official sunspot number series between 1853 and 1858. The {\bf modern re-count} presented here {\bf follows a distribution very similar to the sunspot counts recorded in the Mittheilungen} \citep{1865MiZur...2..193W} for the above-cited two years ({\bf 1859-1860}). In addition, {\bf modern counts also provide the number of spots for the days when only areas are reported in the Mittheilungen} \citep{1874MiZur...4..173W}. \\

Thanks to the Source books recovered and recently digitized by T. Friedli, we can also {\bf compare Wolf's own counts to the modern re-count of Carrington data} (we can separate Wolf's data from other observers). As expected, Wolf's group and spot counts with the 4-foot telescope (SM) are very close (within 10 to 20\% max.) to the counts made from Carrington data by Wolf and the current re-count by Teague.\\

The comparison of Carrington's original counts (CV2014) with Wolf's portable telescope (PM) also {\bf confirms the fact that Carrington's observations tabulated in the Mittheilungen are in fact a re-count by Wolf}. CV2014 observations almost match Wolf's portable telescope scale, however, while a factor of 1.5 had been applied to the former, Carrington's k factor remained 1.03 and is not coherent with the use of original counts as presented in CV2014. \\

In addition, {\bf comparison of Teague's re-count of Carrington's original drawings with Schwabe's re-counted data} by \cite{2013MNRAS.433.3165A} gives a comparable distribution on the overlapping days while CV2014 shows a large deviation from Teague's data. This {\bf confirms the validation of the re-count by Teague} (2020) as an alternative to the existing data from the Mittheilungen or CV2014 which both lack detailed spot information.\\

Although the current re-count series seems to suffer from a scale discrepancy of $\approx$ 30\% in 1858, a stability study according to the method presented in \cite{2019ApJ...886....7M} indicates that this {\bf modern re-count is stable within its own error bars}. This apparent jump can be attributed to the start of a new cycle and therefore, suffers an increase of sunspot counts as expected. Nonetheless, using the re-count by Teague will be a convenient option for future studies involving Carrington's data.\\

The raw {\bf Schwabe data} which is tabulated in the Mittheilungen, although very interesting {\bf has not been used here} because of a scale issue recently uncovered and linked to the problem already pointed out in \cite{2016SoPh..291.2505F}. The discrepancies and their impact on the original construction of SN need to be understood in order to include raw observations in the reconstruction of SN. \\

We recall that the {\bf overarching goal of this work is the complete reconstruction of the Sunspot Number series from the raw data}. The verification of Teague's re-counted data from Carrington's original drawings as a stable and homogeneous series which is coherent with the other contemporary observers, gives a {\bf reliable set of observations for the period 1853-1861}. \\ 

\begin{acks}
This work was supported by a PhD grant awarded by the Royal Observatory of Belgium to S. Bhattacharya. L. Lef\`evre and F. Clette wish to acknowledge the support of ISSI \url{https://www.issibern.ch/teams/sunspotnoser/}.
\end{acks}
\bibliographystyle{spr-mp-sola}
\bibliography{bilbio}  
\end{article} 
\end{document}